\documentclass{ws-ijmpe}

\begin{document}

\markboth{Yoshimasa Hidaka}{Spectrum of soft mode with thermal mass of 
quarks above critical temperature}

\catchline{}{}{}{}{}

\title{SPECTRUM OF SOFT MODE WITH THERMAL MASS OF QUARKS ABOVE CRITICAL TEMPERATURE}

\author{\footnotesize YOSHIMASA HIDAKA}

\address{RIKEN BNL Research Center, Brookhaven National
 Laboratory, Upton, New York 11973, USA\\
hidaka@quark.phy.bnl.gov}

\author{MASAKIYO KITAZAWA}

\address{RIKEN BNL Research Center, Brookhaven National
 Laboratory, Upton, New York 11973, USA\\
kitazawa@quark.phy.bnl.gov}

\maketitle


\begin{abstract}
We study effects of a thermal-quark mass and Landau damping of quarks on 
the chiral phase transition and its soft modes at finite temperature.
For this purpose we employ a simple model with the quark propagator obtained 
in the hard thermal loop approximation.
We show that the chiral phase transition is second order even 
if quark have a finite thermal mass, and the thermal mass suppresses the 
 chiral condensate.
We argue that mesonic spectra have a large width due to scattering 
 between gluons and quarks, and the van Hove singularity at threshold \cite{Hidaka}.
\end{abstract}

\section{Introduction}
Chiral and confinement-deconfinement phase transitions are the most remarkable phenomena of quantum 
chromodynamics (QCD) in hot and/or dense matter.
The data of the Relativistic Heavy Ion Collider (RHIC) experiment show strongly coupled 
quark and gluon matter near the critical temperature, $T_\text{c}$ \cite{RHIC}.
To explore how mesonic excitations change is interesting problem above $T_\text{c}$.
Recent lattice studies predict that heavy quarkonia such as $J/\Psi$ 
will survive above $T_\text{c}$ \cite{Lattice}.
In the present work we focus on light meson sectors, especially scalar and pseudo-scalar channels.
It is suggested that light corrective modes appear as soft modes of the chiral restoration
as temperature approach to the critical temperature \cite{HK85}.

It is expected that spectra of quarks and gluons have peculiar 
structures above $T_\text{c}$.
In particular at extremely high temperature, it is known that the 
thermal gluon fields  give quarks mass-gap and
Landau-damping in the space-like region \cite{HTL}.
The aim of our work is to investigate the effects of thermal mass of quarks and 
Landau damping of quarks on the chiral phase transition and mesonic excitations.
For this reason, we adopt a simple model with four Fermi interactions 
and quarks having thermal mass \cite{Hidaka}.
We show that behaviors of the chiral condensate, phase transition and spectral function in 
the scalar and pseudo-scalar channel with the thermal mass.
\section{Quasi quarks}
Before exploring mesonic excitations above $T_\text{c}$, we consider 
thermal excitations of the quarks in the chiral limit.
A retarded propagator of the quarks at finite temperature in the chirally 
symmetric phase is written by
\begin{equation}
S^\text{R}(\omega,\bm{p})=i\int \frac{d\omega'}{2\pi}\frac{1}{\omega-\omega'+i\eta}\left(\text{P}_+\rho_+(\omega',\bm{p})+\text{P}_-\rho_-(\omega',\bm{p})\right),
\end{equation}
where $\text{P}_\pm$ are projection operators and $\rho_\pm$ are spectral 
function of quarks. The propagator is chirally symmetric because it anticommute with $\gamma_5$.
In medium the quark spectrum has a value in not only time-like region but also space-like 
region because of scatterings and decays between the quark and gluons,
and have a mass-gap called {\it thermal mass} in spite of chiral 
symmetric one.
In the high temperature limit, the spectral function can be calculated with the HTL 
approximation \cite{HTL},
\begin{align}
\rho _ \pm^\text{HTL}  (\omega ,p) =&-2\text{Im} S^\text{HTL}_{\pm}(\omega,\bm{p})\nonumber\\
=& 2\pi [Z_\pm(p)\delta (\omega  - \omega _ \pm  (p)) + Z_\mp(p)\delta (\omega  + \omega _ \mp  (p))
 + \rho^L _ \pm  (\omega ,p)\theta (p^2  - \omega ^2 )],
\label{eq:quarkspectral}
\end{align}
where $p=|\bm{p}|$, $\omega_+(p)$ and $\omega_-(p)$ correspond to normal quasiquark and plasmino 
excitations respectively and their residues are $Z_{\pm}(p)=(\omega_{\pm}^2-p^2)/(2m_T^2)$.
At $p=0$ normal quasiquark and plasmino excitations are degenerated and 
have the same mass, 
$\omega_\pm(0)=m_T$.
$\rho^L _ \pm  (\omega ,p)$ are continuum spectra in space-like region, which
originate from Landau damping of quarks,
\begin{align}
\rho^L _ \pm  (\omega ,p) = \frac{\frac{m_T ^2}
{{2p^2 }}(p \mp \omega ) }{ {\left( {(p \mp \omega ) \pm \frac{{m_T ^2 }}
{{2p^2 }}\left[ {(p \mp \omega )\ln \left| {\frac{{\omega  + p}}
{{\omega  - p}}} \right| \pm 2p} \right]} \right)^2  + \frac{{\pi ^2 m_T ^2 }}
{{4p^4 }}(p \mp \omega )^2 } }.
\end{align}

Now, we would like to explore the effects of the thermal mass on the chiral 
transition and mesonic excitations.
We consider a following simple model with the quark propagator obtained with the HTL approximation and four Fermi
interactions,
\begin{align}
\mathcal{L} 
= \bar \psi [{S}^\text{HTL}_{m_T}]^{-1} \psi  
+ G_\text{S}[(\bar \psi \psi )^2  + (\bar \psi i\gamma _5 \psi )^2 ],
\label{eq:L}
\end{align}
where $\psi$ is the quark field in the chiral limit,
$G_\text{S}$ is the scalar coupling, and ${S}^\text{HTL}_{m_T}$ is 
the propagator obtained in the HTL approximation.
$m_T$ is introduced as a parameter to be varied by hand in our model.
We introduce the three-dimensional cutoff $\varLambda$ to eliminate
the ultraviolet divergence.
When  $m_T=0$ this model  is equivalent to the Nambu--Jona-Lasinio model.
\section{Chiral restoration with thermal mass}
\begin{figure}[th]
\begin{center}
\includegraphics[width=.65\textwidth]{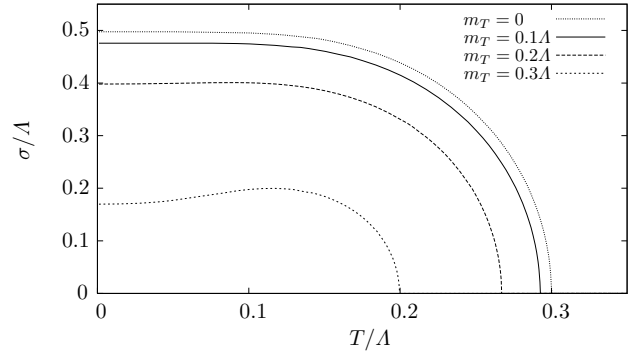}
\end{center}
\caption{
Order parameters for the chiral phase transition
$ \sigma \equiv -2G_\text{S} \langle \bar\psi\psi \rangle $
for $m_T/\varLambda = 0 , 0.1 , 0.2$ and $0.3$ 
with fixed $G_\text{S} \varLambda^2 = 13.01$.
The chiral transition is of second order irrespective of $m_T$.
For larger $m_T$, $\sigma$ becomes smaller.
}
\label{fig:condensate}
\end{figure}

First, we consider behaviors of the chiral phase transition in the mean-field approximation.
In this approximation free energy is given by
\begin{align}
V(\sigma) 
=& 
\frac{ \sigma^2 }{ 4G_\text{S} } - T\sum\limits_n 
\int \frac{ d^3 k} {(2\pi )^3 }
{\text{tr}}\ln \left([{S}^\text{HTL}_{m_T}( i\omega_n , \bm{k})]^{-1}-\sigma\right),
\label{eq:effpot}
\end{align}
where $\omega_n=(2n+1)\pi T$ is Matsubara frequency for fermions. 
$\sigma\equiv-2G_\text{S}\langle\bar{\psi}\psi\rangle$ is the chiral 
condensate which is obtained by minimization of the effective potential,
\begin{align}
\frac{\partial V(\sigma)}{\partial\sigma}=0.
\end{align}
The numerical results of the chiral condensate are shown in Fig.~\ref{fig:condensate}.
Although thermal-quark mass generally depends on $T$,
we fix the thermal mass to $m_T/\varLambda=0$, $0.1$, $0.2$ and 
$0.3$ in order to study properties of thermal effects. There are two interesting 
features in Fig.~\ref{fig:condensate}: 
First, the chiral condensate becomes small as the thermal mass increases.
The reason is that additional pairing energy is required to pair the quark
and anti-quark when quarks have a mass gap.
This feature is completely different from the effect of Dirac mass on 
the chiral condensate. 
Because the Dirac mass act as an external field breaking chiral 
symmetry, it increases the chiral condensate like an external source
in a spin system with  although Dirac masses play a role of mass gap
as well as thermal masses. 
Second, at $T_\text{c}$ in Fig.~\ref{fig:condensate} the chiral 
condensate continuously becomes zero, i.e. the chiral transition is second order.
Thermal mass does not change the order of the chiral phase transition
because it does not break chiral symmetry unlike Dirac mass. 
\section{Mesonic spectra in scalar and pseudoscalar channel}
\begin{figure}[t]
\begin{center}
\begin{align}
D^\text{R}_\sigma =& G_\text{S} + 
\parbox{12mm}{\includegraphics[width=12mm]{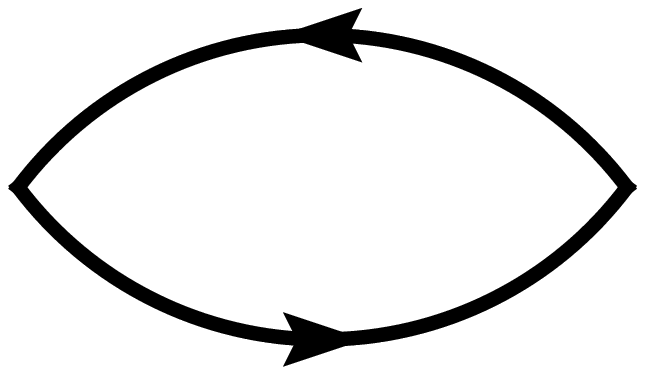}} + 
\parbox{24mm}{\includegraphics[width=12mm]{polarization2.eps}\hspace{-0.1cm}
\includegraphics[width=12mm]{polarization2.eps}} + \cdots 
= \frac{-1}{ G_\text{S}^{-1} + \varPi^\text{R}_\sigma }
\nonumber \\
\varPi^\text{R}_\sigma = &
\parbox{12mm}{\includegraphics[width=12mm]{polarization2.eps}} 
=
\parbox{15mm}{\includegraphics[width=15mm]{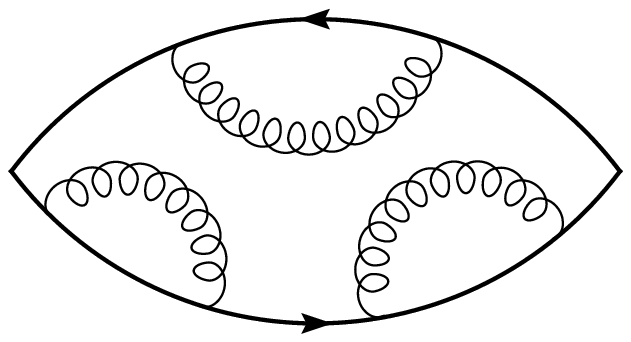}} + \cdots,
\hspace{4mm}
\parbox{12mm}{\includegraphics[width=12mm]{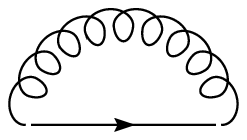}} 
= \varSigma^{\rm HTL}_{m_T}
\nonumber
\end{align}
\caption{
Diagrammatic representation for the meson propagator 
$D^\text{R}_\sigma(\omega,\bm{p})$ and the polarization function 
$\varPi^\text{R}_\sigma(\omega,\bm{p})$.
The thick-solid line denotes the quark propagator
in the HTL approximation Eq.~(\ref{eq:quarkspectral}), while
the thin-solid line represents the free quark propagator.
}
\label{fig:resum}
\end{center}
\end{figure}
\begin{figure} 
\begin{tabular}{cccc}
\parbox{.5cm}{(a)\vspace{.9cm}} &
\parbox{2cm}{\includegraphics[width=.17\textwidth]{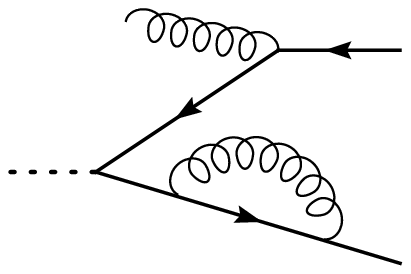}} & \hspace{25mm}
\parbox{.5cm}{(b)\vspace{.9cm}} &
\parbox{2cm}{\includegraphics[width=.17\textwidth]{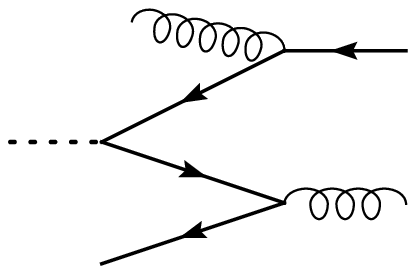}}
\end{tabular}
\caption{
The diagrammatic al interpretation for the decay processes 
included in $I_\text{PC}(\omega)$ (left) and $I_\text{CC}(\omega)$ (right).
The dashed line denotes the mesonic excitations and
the temporal direction is left to right.
}
\label{fig:diagram}
\end{figure}
\begin{figure}[tbp]
\begin{center}
\includegraphics[width=.55\textwidth]{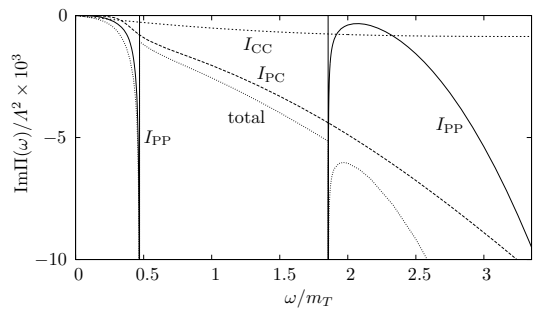}
\caption{
Each component of Im$\varPi^\text{R}(\omega,\bm{p}=\bm{0}) 
= I_{\rm PP}(\omega) + I_{\rm PC}(\omega) + I_{\rm CC}(\omega)$
at $T/\varLambda = m_T/\varLambda = 0.3$.
$I_{\rm PP}(\omega)$, $I_{\rm PC}(\omega)$ and $I_{\rm CC}(\omega)$ 
include the decay processes into
two quasiparticle poles, a quasiparticle pole and continuum, and
two continuum, respectively.
}
\label{fig:imag}
\end{center}
\end{figure}

Near $T_\text{c}$ soft-modes are expected to appear as fluctuations of order 
parameter even if the quark has a thermal mass because the thermal mass 
does not break chiral symmetry above $T_\text{c}$ \cite{HK85}.
The scalar and pseudo-scalar modes are degenerated above $T_\text{c}$, 
thus we consider only scalar mode in the following.
In the random phase approximation, the retarded propagator of scalar channel, 
$D^\text{R}_\sigma(\omega, \bm{p})$, is given by
\begin{align}
D_\sigma^\text{R}(\omega,\bm{p})=\frac{-1}{G_\text{S}^{-1}+\varPi^\text{R}_\sigma(\omega,\bm{p})},
\end{align}
where $\varPi^\text{R}_\sigma(\omega,\bm{p})$ is the 
one-loop self-energy in the scalar channel, corresponding to the second diagram in Fig.~\ref{fig:resum}.
$\varPi^\text{R}_\sigma(\omega,\bm{p})$ is obtained by analytical continuation, $\varPi^\text{R}_\sigma(\omega,\bm{p})=\tilde{\varPi}_\sigma(i\nu_n,\bm{p})|_{i\nu_n\to\omega+i\eta}$
in the imaginary formalism, where $\nu_n=2\pi n$ is Matsubara frequency.
$\tilde{\varPi}_\sigma(i\nu_n,\bm{p})$ is given by
\begin{eqnarray}
\tilde\varPi_\sigma( i\nu_n,\bm{p} )
&=&
2T\sum_m\int\frac{ d^3k}{(2\pi)^3}
{\text{tr}} [
S^\text{HTL}_{m_T}(i\omega_m,\bm{k})
S^\text{HTL}_{m_T}(i\nu_n+i\omega_m,\bm{p}+\bm{k}) ].
\label{eq:Pi_def}
\label{eq:Pi}
\end{eqnarray}
Let us explore the imaginary part of $\varPi^\text{R}_\sigma$ which corresponds to decays and 
scatterings of the mesonic excitations with thermal particles in the heat bath.
At $p=0$ there are the following three types of contributions:
\begin{align}
{\rm Im} \varPi^\text{R}_\sigma( \omega,\bm{0} )
= -\frac 1\pi (
I_\text{PP}(\omega) + I_\text{PC}(\omega) + I_\text{CC}(\omega) ),
\end{align}
with
\begin{align}
I_\text{PP}( \omega )
=& 
- \frac{ p^2 Z_+^2 }{ 2 |\omega_+'| } [ 1-2f_+ ]
\big|_{\omega=-2\omega_+ }
- \frac{ p^2 Z_-^2 }{ 2 |\omega_-'| } [ 1-2f_- ]
\big|_{\omega=2\omega_- }\nonumber\\
&- \frac{ p^2 Z_+ Z_- }{ |\omega_+'  - \omega_-'| }
[ f_+ - f_- ]
\big|_{\omega = \omega_- - \omega_+ } -(\omega\leftrightarrow-\omega),
\label{eq:imPP}
\\
I_\text{PC}( \omega )
=& 
2\pi\int \frac{d^3k}{(2\pi)^3} Z_+ [ f(\omega_+) - f(\omega+\omega_+) ] 
\rho_-^\text{L} ( \omega+\omega_+ ,\bm{k} ) \nonumber\\
&+2\pi\int \frac{d^3k}{(2\pi)^3} Z_- [ f(\omega_- -\omega) - f(\omega_-) ] 
\rho_+^\text{L} (  \omega - \omega_-  ,\bm{k} )
-(\omega\leftrightarrow-\omega),
\label{eq:imPC}
\\
I_\text{CC}(\omega)
=&
\frac{1}{2}\int \frac{d^3k}{(2\pi)^3} \int dE [ f(E) - f(\omega+E) ]
\rho_+^\text{L}(E,\bm{k}) \rho_-^\text{L} ( \omega+E, \bm{k} ) 
-(\omega\leftrightarrow-\omega),
\label{eq:imCC}
\end{align}
where indices of PP, PC and CC correspond to the contribution of pole-pole, 
pole-continuum and continuum-continuum respectively, and 
$f(E)=[\exp(E/T)+1]^{-1}$, $f_{\pm}=f(\omega_{\pm})$ and $\omega_{\pm}'=d\omega(p)/dp$.
Imaginary part of the self-energy for a typical parameter set $T/\varLambda=m_T/\varLambda=0.3$ is shown in Fig.~\ref{fig:imag}.
Contributions in $I_\text{PP}(\omega)$ correspond to decays of the soft-mode
into normal quasiquark and anti-quasiquark (plasmino and anti-plasmino) 
above $1.8m_T$ and correspond to the scattering between 
the soft-mode and the plasmino below $0.4m_T$.
$I_\text{PP}(\omega)$ has divergences at two thresholds, 
$\omega\simeq0.4m_T$ and $1.5m_T$. This singularities are called van Hove singularities,
and come from divergences of the density of states with $2\omega_-(p)$ 
and $\omega_-(p)-\omega_+(p)$.
On the other hand $I_\text{PC}$ and $I_\text{CC}$ are continuous 
functions and have finite values except for $\omega=0$.
Typical processes in $I_\text{PC}$ and $I_\text{CC}$ are shown in Fig.~\ref{fig:diagram}.
$I_\text{PC}$ have large values, this contributions are actually larger 
than spectral function with $m_T=0$.

\begin{figure}[tbp]
\begin{center}
\includegraphics[width=.55\textwidth]{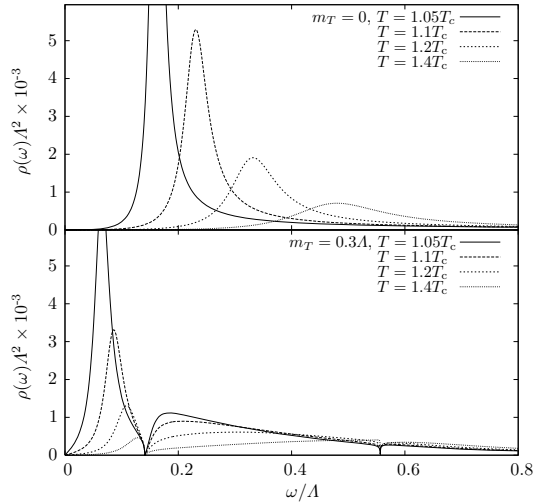}
\end{center}
\caption{
Spectral functions for $m_T=0$ (upper) and $m_T/\varLambda=0.3$ (lower)
with several values of $T$ above $T_c$. 
The coupling $G_\text{S}$ is determined so that $T_\text{c}/\varLambda=0.3$
for each $m_T$.}
\label{fig:spectrum}
\end{figure}

In Fig.~\ref{fig:spectrum} we show spectral functions in scalar channel
for $m_T/\varLambda=0$ and $0.3$ at $\bm{p}=\bm{0}$
in the upper and lower panels, respectively, and the scalar coupling $G_\text{S}$ is adjusted
so that $T_\text{c}/\varLambda = 0.3$ for each $m_T$.
In the upper panel with $m_T=0$, there appears a peak in $\rho_\sigma(\omega,\bm{p})$, 
and the peak becomes sharper and moves toward the origin
as $T$ approaches $T_\text{c}$ from high temperature.
This is nothing but the soft mode of the chiral phase transition \cite{HK85}.
In the lower panel with finite $m_T$, 
$\rho_\sigma(\omega,\bm{p})$ behaves in a more complicated way.
First, 
$\rho_\sigma(\omega,\bm{p})$ becomes zero at two energies
$\omega_1$ and $\omega_2$ corresponding to van Hove singularities
in $I_\text{PP}(\omega)$.
Second, 
although there appears a sharp peak in $\rho_\sigma(\omega,\bm{p})$
as the soft mode of the chiral transition, 
the heights of the peaks in $\rho_\sigma(\omega,\bm{p})$ 
are depressed compared from the upper panel with $m_T=0$ because of 
large scattering between the quasi-quarks and gluons.

\section{Summary}
We studied the effects of thermal mass and Landau 
damping of quarks on the chiral phase transition and mesonic spectra at 
finite temperature.
The thermal mass does not break chiral symmetry so that the order of 
chiral phase transition does not change.
On the other hand, the thermal mass makes the chiral condensate small.
In mesonic spectra not only thermal mass but also Landau-damping of 
quarks play important role.
Both thermal mass and Landau damping are specific at finite temperature, 
and they are independent from mesonic channels.
Hence our result suggests that the mesonic excitations in the light quark sector are destroyed by the thermal effects
above $T_\text{c}$.
It is notable that there can nevertheless appear sharp 
peaks in the mesonic spectra as the soft modes of the chiral transition 
near $T_\text{c}$ \cite{HK85}.

\end{document}